\pdfoutput=1
\documentclass[10pt,twocolumn]{article}
\usepackage[a4paper, margin=1.5cm, bmargin=2cm]{geometry}
\usepackage{indentfirst}
\usepackage[utf8]{inputenc}
\usepackage[english]{babel}
\usepackage{textgreek}
\usepackage{amsmath}
\usepackage{amsfonts}
\usepackage{amssymb}
\usepackage[per-mode=symbol]{siunitx}
\usepackage{bm}                       % enhanced bold in math mode
\usepackage{mathpazo}                 % pretty font
\usepackage{microtype}                % better kerning
\usepackage{tikz}
\usepackage{graphicx}
\usepackage{float}                    % to use float placement H
\usepackage[export]{adjustbox}        % to center wide figure
\usepackage[keeplastbox]{flushend}    % balance columns on last page
\usepackage[font=small,labelfont=bf]{caption}
\usepackage{subfig}
\usepackage{ifthen}                   % for if condition in newcommand
\usepackage{booktabs}                 % pretty tables
\usepackage{multirow}                 % fuse cells in tables
\usepackage[symbol]{footmisc}         % symbols for footnotes

\usepackage{hyperref}
\hypersetup{colorlinks=true, linkcolor=blue, citecolor=blue, urlcolor=blue} % NOT SUITED FOR PRINTING!

\usepackage{xcolor}
\definecolor{blue1}{RGB}{ 7,  47,  95}
\definecolor{blue2}{RGB}{18,  97, 160}
\definecolor{blue3}{RGB}{56, 149, 211}
\definecolor{red}{RGB}{210, 0, 0}

\usepackage{titlesec}                 % section customisation
\titlelabel{\thetitle.\ }             % period after section numbers
\titleformat*{\section}{\color{blue1}\scshape\bfseries\centering\large}
\titleformat*{\subsection}{\color{blue2}\normalfont\itshape\large}
\titleformat*{\subsubsection}{\color{blue3}\normalfont\itshape}
\titleformat{\paragraph}[runin]{\color{blue3}\normalfont\itshape}{}{0em}{}[~-]
\titlespacing{\paragraph}{0em}{0em}{0.3em}

\usepackage[backend=bibtex,date=year,doi=false,giveninits=true,intitle=true,isbn=false,maxnames=1,maxbibnames=99,sorting=none,style=nature]{biblatex}
\AtEveryBibitem{
	\clearlist{language}
}
\newlength{\spc} % declare a variable to save spacing value
\let\footnoteorig\footnote
\renewcommand{\footnote}[2]{% #1: footnote text, #2: punctuation
	\ifthenelse{\equal{#2}{,}\OR\equal{#2}{.}}{%
		\settowidth{\spc}{#2}% set value of \spc variable to the width of #2 argument
		\addtolength{\spc}{-1.8\spc}% subtract from \spc about two (1.8) of its values making its magnitude negative
		#2% print the punctuation
		\hspace*{\spc}% print an additional negative spacing stored in \spc after #2
		\footnoteorig{#1}% print the superscript number
	}{%
		\footnoteorig{#1}%
		\ifthenelse{\NOT\equal{#2}{;}\AND\NOT\equal{#2}{:}}{\ }{}%
		#2%
	}%
} % USE sidenoteorig IN SECTIONS!!

\renewcommand{\textcite}[1]{\citeauthor{#1}\hspace*{-0.15em}\supercite{#1}}
\renewcommand{\cite}[2]{% #1: citation string, #2: punctuation
	\ifthenelse{\equal{#2}{,}\OR\equal{#2}{.}}{%
		\settowidth{\spc}{#2}% set value of \spc variable to the width of #2 argument
		\addtolength{\spc}{-1.8\spc}% subtract from \spc about two (1.8) of its values making its magnitude negative
		#2% print the punctuation
		\hspace*{\spc}% print an additional negative spacing stored in \spc after #2
		\supercite{#1}% print (cite) the citation
	}{%
		\supercite{#1}%
		\ifthenelse{\NOT\equal{#2}{;}\AND\NOT\equal{#2}{:}}{\ }{}%
		#2%
	}%
}

\newcommand{\snspace}[2][0.45em]{% #1: punctuation
	\hspace*{-#1}#2\hspace{0.2em}}

\begin{document}

\twocolumn[
	\begin{@twocolumnfalse}

		\begin{center}
			\textbf{\color{blue1}\Large Creation and evolution of roughness on silica under unlubricated wear}\\
			\vspace{1em}
			Son Pham-Ba\footnotemark\hspace*{-0.35em},\hspace{0.1em} Jean-François Molinari\\\vspace{0.5em}
			\textit{\footnotesize Institute of Civil Engineering, Institute of Materials Science and Engineering,\\\vspace{-0.2em}
			École polytechnique fédérale de Lausanne (EPFL), CH 1015 Lausanne, Switzerland}
		\end{center}

		% \vspace{0.5em}

		\begin{center}
			% \textit{Abstract}\par
			% \vspace{0.5em}
			\parbox{14cm}{\small
				\setlength\parindent{1em}Friction and wear are important phenomena occurring in all devices with moving parts. While their origin and the way they evolve over time are not fully understood, they are both intimately linked to surface roughness. Guided by \emph{pin-on-disc} experiments, we present the steps giving rise to the formation of surface roughness on silica, first by the creation of roughly spherical wear particles whose size is related to a critical length scale governing the transition between ductile and brittle behavior, then by the accumulation of these small particles into a larger third body layer, or gouge. We show that, for the explored range of loading conditions, the surface roughness evolves toward a common steady state under unlubricated wear regardless of the initial surface roughness, hinting toward the possible predictability of roughness evolution in wearing components. The friction coefficient is shown to be related to the surface roughness and its time evolution is discussed as well.

				\vspace{1em}
			}
		\end{center}

		\vspace{1em}

	\end{@twocolumnfalse}
]

\footnotetext{Corresponding author. E-mail address: \href{mailto:son.phamba@epfl.ch}{son.phamba@epfl.ch}}

\section{Introduction}

Friction coefficients and the way they evolve over time are currently hardly predictable. In order to better understand friction, it is useful to relate it to known geometrical and physical characteristics of the tribological system. As the most simple friction laws suggest, like the Amontons-Coulomb's friction law\cite{amontonsResistanceCauseeDans1699,coulombTheorieMachinesSimples1785}, friction forces are mostly independent of the apparent contact area, and are instead linked to the real contact area\cite{bowdenFrictionLubricationWear1966}. Indeed, contacting rough surfaces only touch where protruding asperities meet, resulting in a real contact area being only a fraction of the total apparent contact area\cite{greenwoodContactNominallyFlat1966,bushElasticContactRough1975,perssonElastoplasticContactRandomly2001,hyunFiniteelementAnalysisContact2004} in most operating conditions. The ratio between real and apparent contact area is positively correlated with the normal load and is smaller for increasing surface roughness. Since frictional processes can only take place at the contact spots in an unlubricated contact, this highlights a direct relation between friction and surface roughness. This relation applies to all engineered\cite{mandelbrotFractalCharacterFracture1984,majumdarFractalCharacterizationSimulation1990} or natural surfaces\cite{thomNanoscaleRoughnessNatural2017}, as they all display some roughness within a certain range of length scales.

Surface roughness itself can evolve thanks to wear. Molecular dynamics simulations of adhesive wear\cite{milaneseEmergenceSelfaffineSurfaces2019a} show that, at the nanoscale, two rough surfaces sliding on each other evolve toward fractal self-affine rough surfaces with the same fractal dimension regardless of the initial roughness. Another feature of these nanoscale simulations is the apparition of a rolling third-body particle, which seems to be a key factor behind the evolution of the surface roughness to self-affine characteristics.

At the engineering scale, experiments show that wear leads to the modification of the rubbed surfaces, with the formation of a third-body layer made of sintered worn material\cite{wirthFundamentalTribochemicalStudy1994,meierhoferThirdBodyLayer2014}, also called tribo-layer\cite{riahiRoleTribolayersSliding2001} or tribofilm\cite{gosvamiMechanismsAntiwearTribofilm2015,katoTribofilmFormationMild2007}, which contributes to a change in the frictional properties of the interface.

The same evolutionary behavior is found at the geological scale. Gouge formation is omnipresent in brittle faults\cite{scholzWearGougeFormation1987,wilsonParticleSizeEnergetics2005}. It is made of crushed and worn rock particles and dictates the frictional properties of the interface, as seen experimentally\cite{starfieldMethodologyRockMechanics1988,biegelFrictionalPropertiesSimulated1989,mairInfluenceGrainCharacteristics2002} and numerically\cite{morganNumericalSimulationsGranular1999,guoFaultGougeEvolution2007}.

Studies also considered the evolution of surface roughness due to wear, for example in engineering applications using \emph{pin-on-disc} setups, where metallic samples can go through multiple wear regimes over time\cite{kubiakSurfaceMorphologyEngineering2011} and settle down to a steady state after a certain amount of time. At steady state, the worn surfaces of the samples are seen to exhibit identical geometrical characteristics (wear track and surface roughness) regardless of the initial surface roughness for the same operating conditions\cite{yuanSurfaceRoughnessEvolutions2008}. Other studies reproducing or looking at geological phenomena observe a steady state self-affine roughness upon sliding after a running-in period\cite{powerRoughnessWearBrittle1988,brodskyFaultsSmoothGradually2011,candelaMinimumScaleGrooving2016,thomNanoscaleRoughnessNatural2017}. However, overall, the mechanisms of the onset of wear and their link with following wear processes are not studied in detail. A direct non-empirical relation between material properties and wear processes remains to be found.

Here, an experimental study of friction and wear was performed on amorphous silica (SiO$_2$) using a rotating tribometer in a \emph{pin-on-disc} configuration. From observations of worn surfaces after various sliding distances, we deduced the steps giving rise to the creation of a surface roughness from comparatively flat initial surfaces --- from the formation of the first third-body particles, often overlooked in the literature, to the full third-body layer --- while attempting to directly relate the very first events of wear to material properties. The observations linked to the onset of wear are complemented by molecular dynamics (MD) simulations. We also assessed the evolution of the friction coefficient and of the surface roughness given different initial surface roughnesses, and we provide a mechanistic description of the factors behind surface roughness evolution.

\section{Experimental setup}

\begin{figure}[t]
	\includegraphics{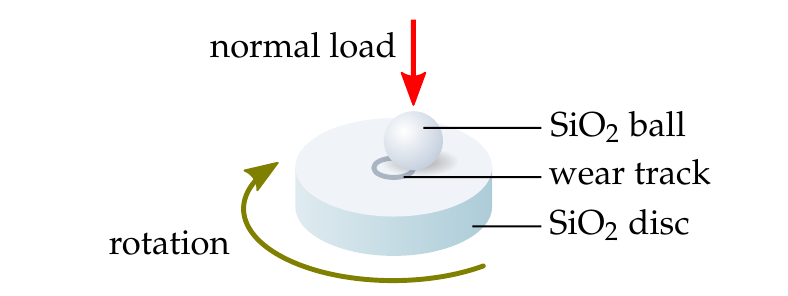}
	\caption[Experimental setup on tribometer]{Experimental setup on tribometer. The disc is fixed to a rotating motor. The ball can only move vertically and is kept against the disc with a constant normal load.}
	\label{fig:tribo}
\end{figure}

The tribological tests have been performed using \SI{6}{mm} spheres of amorphous SiO$_2$ sliding (rotating) on discs made of the same material (see Figure~\ref{fig:tribo}). A \emph{Bruker UMT 3} tribometer was used. The sliding velocity of the ball was \SI{8}{\mm\per\s} and it followed a circular path of radius \SI{1}{mm} with a normal load of $F_\text{N} = \SI{1}{N}$. Using Hertz contact theory, we get that in the elastic case the diameter of the contact zone between the unworn ball and the disc is \SI{78.2}{\micro\m} and the maximum pressure (reached at the center of the contact zone) is \SI{312}{MPa} (see Appendix~\ref{apx:hertz}).

The initial untreated surface roughness of the balls and the discs has been identified using atomic force microscopy (AFM) and is around $Sa \approx \SI{10}{nm}$, which is very small from an engineering point of view (see Appendix~\ref{apx:Sa} for details about the computation of the $Sa$ surface roughness). Some discs have been manually polished using P120 and P80 sandpaper, resulting in roughened surfaces with $Sa = \SI{0.74}{\micro\m}$ and $Sa = \SI{1.41}{\micro\m}$ in average respectively (computed from topographic images of size $\SI{117}{\micro\m} \times \SI{88}{\micro\m}$ and resolution $\SI{1360}{px} \times \SI{1024}{px}$ obtained on a \emph{Sensofar S-Neox} confocal microscope). Only the discs were treated to have an initial surface roughness, while the balls were kept flat, with their initial roughness of $Sa \approx \SI{10}{nm}$. It is indeed difficult to polish the SiO$_2$ balls in the same way as the discs without disrupting their spherical shape. Shot peening or sandblasting could have been used to roughen the balls, but having them initially smooth is not very important for this study since a surface roughness (dissimilar to the one on the discs) eventually develops on them.

Table~\ref{tab:exp_rough} lists the tests that were performed. The initial $Sa$ of the discs are indicated. Some variation in the $Sa$ roughnesses can be seen amongst the discs which were polished with the same grain size. This is in part due to the inherent variability of the manual polishing process. The fact that those roughness measurements were only performed in a single small window on each disc can also imply some uncertainty.

\begin{table}
	\caption[Duration and initial roughness of the tests]{Duration and initial roughness of the tests. The table is subdivided into three sections according to the polishing of the discs. The latter are manually polished using sandpaper with the indicated grain sizes. Each test has a duration and a corresponding sliding distance, known from the fixed sliding velocity of \SI{8}{\mm\per\s}. The third column indicates the number of repetitions for each set of parameters, and the multiple initial disc $Sa$ are indicated for each repetition when they differ significantly.}
	\label{tab:exp_rough}
	\centering
	\vspace{0.2cm}
	\footnotesize
\begin{tabular}{ccccc}
	\toprule
	\textbf{Duration} & \textbf{Distance} & \textbf{Rep.} & \textbf{Polishing} & \textbf{Disc} $\bm{Sa}$ [\si{\micro\m}] \\
	\midrule
	\SI{1}{s}    & \SI{8}{mm}   & 1 & \multirow{9}{*}{none} & \multirow{9}{*}{$\approx 0.01$} \\
	\SI{15}{s}   & \SI{12}{cm}  & 1 & & \\
	\SI{30}{s}   & \SI{24}{cm}  & 1 & & \\
	\SI{1}{min}  & \SI{48}{cm}  & 1 & & \\
	\SI{15}{min} & \SI{7.2}{m}  & 1 & & \\
	\SI{30}{min} & \SI{14.4}{m} & 1 & & \\
	\SI{1}{h}    & \SI{28.8}{m} & 3 & & \\
	$2$h$30$     & \SI{72}{m}   & 2 & & \\
	\SI{5}{h}    & \SI{144}{m}  & 3 & & \\
	\midrule
	$2$h$30$     & \SI{72}{m}   & 3 & \multirow{2}{*}{P120} & $0.78$,\quad $0.91$,\quad $0.74$ \\
	\SI{5}{h}    & \SI{144}{m}  & 3 & & $0.79$,\quad $0.63$,\quad $0.58$ \\
	\midrule
	$2$h$30$     & \SI{72}{m}   & 3 & \multirow{2}{*}{P80} & $1.54$,\quad $0.75$,\quad $1.51$ \\
	\SI{5}{h}    & \SI{144}{m}  & 3 & & $1.51$,\quad $1.56$,\quad $1.58$ \\
	\bottomrule
\end{tabular}

	\vspace{0.2cm}
\end{table}

For each test, the $Sa$ roughness of the disc is measured before and after the experiment (details in Appendix~\ref{apx:Sa}). The vertical reaction force $F_z$ and the lateral force $F_x$ parallel to the direction of sliding are recorded during the whole test, from which a dynamical friction coefficient $\mu = F_x/ F_z$ is deduced. Roughness measurements cannot be similarly performed over the whole duration of a test because they are not \emph{in situ} measurements. Finally, scanning electron microscope (SEM) images of the wear tracks and wear particles formed on the initially flat discs were taken after various durations.

\section{Results and discussion}

\subsection{Initial wear particle size}\label{sec:ini_particle}

\begin{figure*}[t]
	\centering
	\subfloat[\label{subfig:sem_1s_a}]{
		\includegraphics{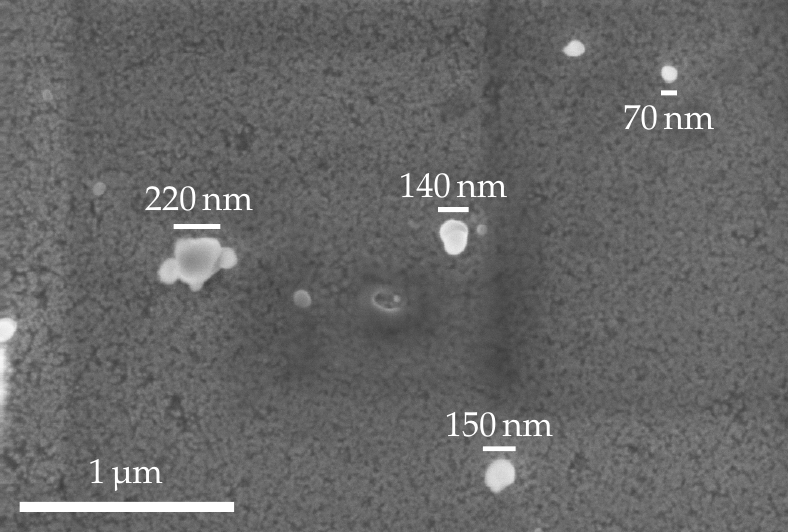}
	}
	\subfloat[\label{subfig:sem_1s_b}]{
		\includegraphics{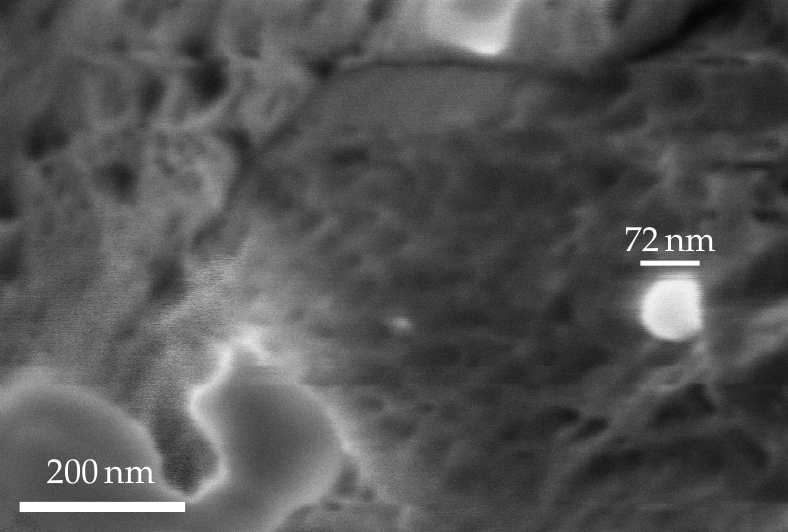}
	}
	\\
	\subfloat[\label{subfig:sem_1s_c}]{
		\includegraphics{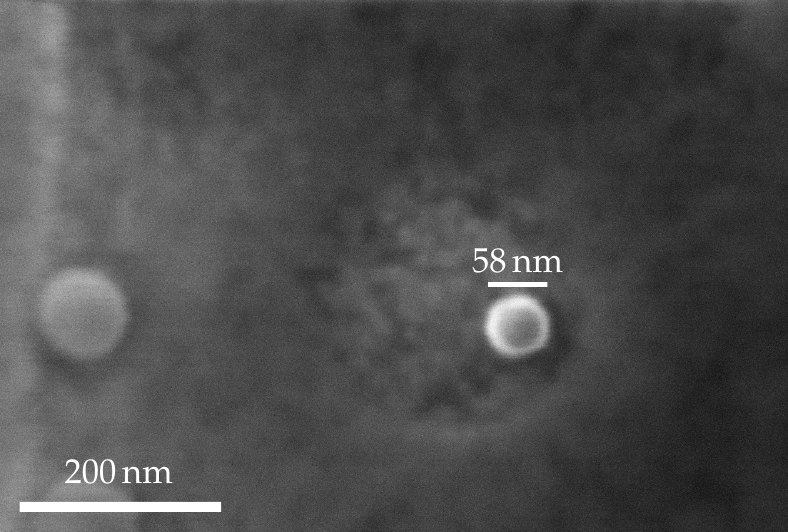}
	}
	\subfloat[The silica was discriminated from the nickel background using energy-dispersive X-ray (EDX) spectroscopy\label{subfig:sem_1s_d}]{
		\includegraphics{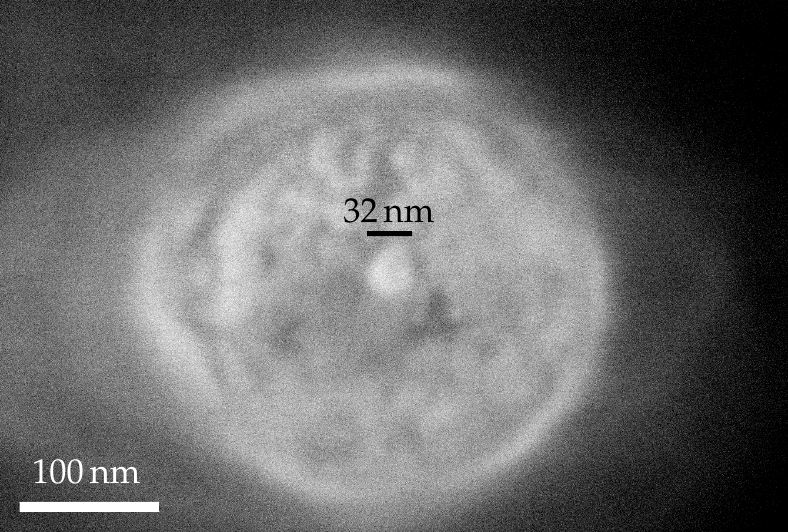}
	}
	\caption[SEM images of wear particles taken from a flat disc after \SI{1}{s} (\SI{8}{mm}) of sliding]{SEM images of wear particles taken from a flat disc after \SI{1}{s} (\SI{8}{mm}) of sliding. The particles were collected from the disc using an adhesive nickel patch for an easier observation.}
	\label{fig:sem_1s}
\end{figure*}

Figure \ref{fig:sem_1s} shows SEM images of wear particles found on a flat disc after only \SI{1}{s} of sliding, corresponding to a sliding distance of \SI{8}{mm}. The images were obtained using a \emph{Zeiss SUPRA 55-VP} SEM. The formation of wear particles is the first and only evidence of wear found on the disc at this stage (there is no wear track). An adhesive nickel patch was rubbed against the surface of the worn disc to capture the wear particles in order to more easily observe them with SEM. No particle smaller than \SI{30}{nm} was observed. Whether this size depends on the material properties or the loading conditions is discussed thereafter.

Recent theoretical work\cite{aghababaeiCriticalLengthScale2016} has confirmed the existence of a critical length scale $d^*$ which governs a transition between ductile and brittle behavior for any given material. The expression of $d^*$ for the detachment of spherical wear particles is
\begin{equation}\label{eq:d_star_ramin}
	d^* = \frac{12\gamma G}{\sigma_\text{j}^2} \,,
\end{equation}
where $\gamma$ is the surface energy of the material, $G$ is the shear modulus and $\sigma_\text{j}$ is the shear strength (or flow stress) of cold welded junctions. Note that $d^*$ is mainly dependent on material parameters and weakly depends on geometry and loading conditions (which influence the `$12$' factor), thus it can be itself considered as a material parameter. A sheared junction between two surfaces can only be detached into a wear particle if its size $d$ is larger than $d^*$\snspace. Otherwise, it flows plastically. It implies that no wear particles smaller than $d^*$ should be observed in wear experiments, because they cannot be created.

Using MD simulations, we computed estimates for the unknown material parameters of amorphous SiO$_2$ using a potential by Vashishta et al\cite{vashishtaInteractionPotentialSiO21990}. taking into account three-body interactions, using a cutoff parameter of $r_\text{c} = \SI{8}{\angstrom}$. The amorphous SiO$_2$ MD system is prepared following the procedure of Luo et al.\cite{luoSizeDependentBrittletoDuctileTransition2016}, starting from a \textbeta-crystobalite structure. The system is heated and cooled by equilibrating it for 90000 timesteps of \SI{0.5}{fs} at 5000, 4000, 3000, 2500, 2000, 1500, 1000, 500 and \SI{300}{K}, with cooling periods with a rate of \SI{166}{K/ps} between each constant temperature step, except when going to \SI{300}{K} where a rate of \SI{13}{K/ps} is used. The pressure is always kept at 0 using a Berendsen barostat. The isotropic material properties of the obtained material at \SI{0}{K} are $E = \SI{120}{GPa}$, $\nu = 0.22$ and $G = \SI{49}{GPa}$. A shear strength of $\sigma_\text{j} = \SI{7.2}{GPa}$ is obtained from a bulk shear test without fracture, with periodic boundary conditions, at a strain rate of \SI{4e7}{s^{-1}}. A surface energy of \SI{1.5}{N/m} is estimated by cutting the sample at 10 random planes and averaging the results, while the charge neutrality is being preserved by keeping periodic boundary conditions. Using \eqref{eq:d_star_ramin} with those material parameters leads to the estimate:
\begin{equation*}
	d^* \approx \SI{15}{nm} \,.
\end{equation*}
Luo et al\cite{luoSizeDependentBrittletoDuctileTransition2016}. observed a similar ductile-to-brittle transition size of \SI{18}{nm} in MD simulations of glass SiO$_2$ nanofibers, also using the Vashishta potiential. It is also shown that this potential can provide quantitatively correct values for Young's modulus and tensile strength, so that the value found for $d^*$ is likely to also be quantitatively accurate.

\begin{figure*}[p]
	\centering
	\subfloat[Initial configuration\label{subfig:md_07_ini}]{
		\includegraphics{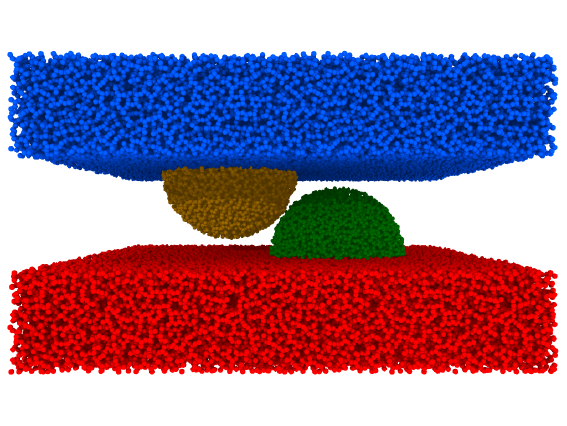}
	}
	\subfloat[After collision\label{subfig:md_07_fin}]{
		\includegraphics{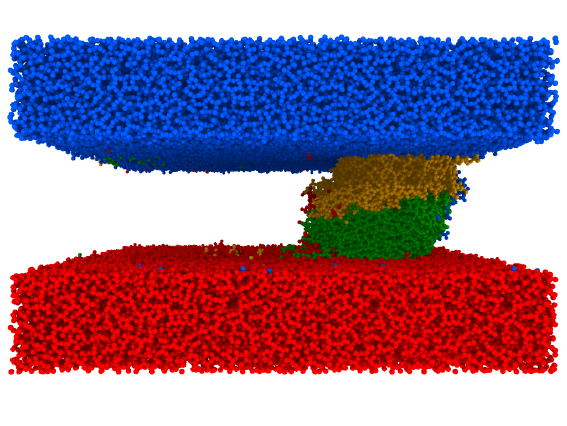}
	}
	\subfloat[Shear strain (cross section)\label{subfig:md_07_def}]{
		\includegraphics{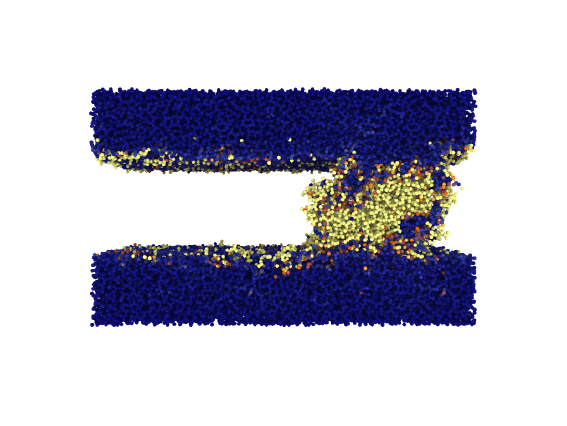}
	}
	\caption[MD simulation of a collision between two amorphous SiO$_2$ asperities of diameter $d = \SI{7}{nm}$]{MD simulation of a collision between two amorphous SiO$_2$ asperities of diameter $d = \SI{7}{nm}$. The initial simulation size is $\SI{20.2}{nm} \times \SI{21.0}{nm} \times \SI{11.8}{nm}$ (in the figures: length $\times$ height $\times$ depth). The non-localized shear strain (brighter is higher) shows a plastic deformation of the asperities, indicating that $d < d^*$.}
	\label{fig:md_07}
\end{figure*}

\begin{figure*}[p]
	\centering
	\subfloat[Initial configuration\label{subfig:md_10_ini}]{
		\includegraphics{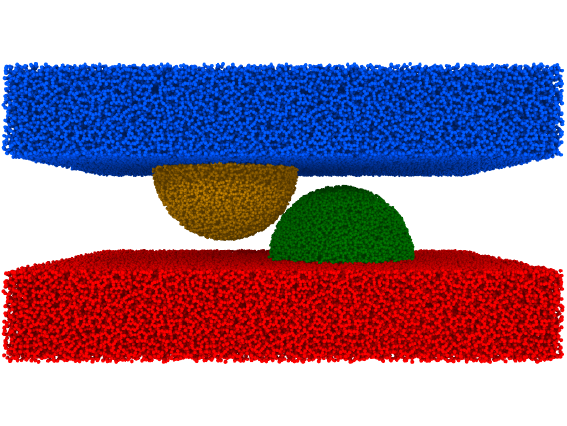}
	}
	\subfloat[After collision\label{subfig:md_10_fin}]{
		\includegraphics{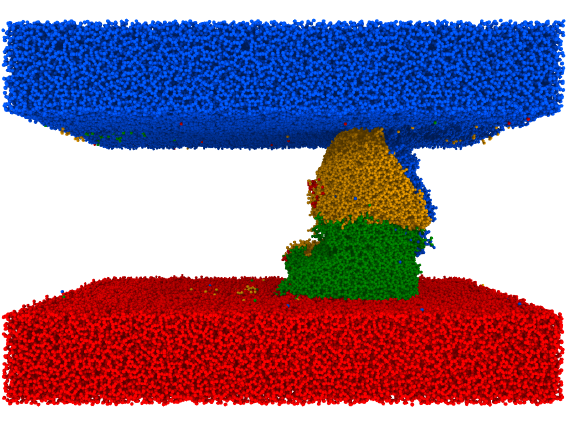}
	}
	\subfloat[Shear strain (cross section)\label{subfig:md_10_def}]{
		\includegraphics{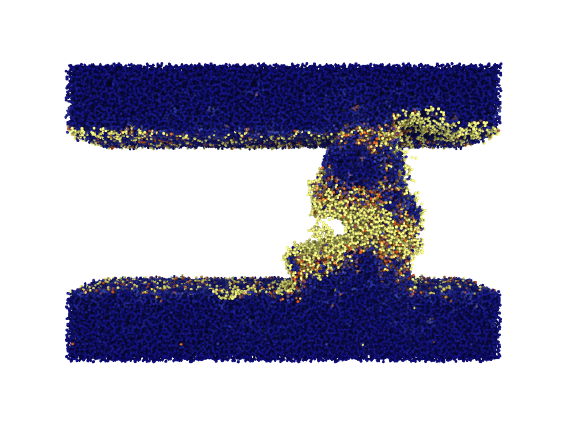}
	}
	\caption[MD simulation of a collision between two amorphous SiO$_2$ asperities of diameter $d = \SI{10}{nm}$]{MD simulation of a collision between two amorphous SiO$_2$ asperities of diameter $d = \SI{10}{nm}$. The initial simulation size is $\SI{30.3}{nm} \times \SI{21.0}{nm} \times \SI{16.0}{nm}$. The behavior is in-between ductile and brittle, indicating that $d \simeq d^*$.}
	\label{fig:md_10}
\end{figure*}

\begin{figure*}[p]
	\centering
	\subfloat[Initial configuration\label{subfig:md_20_ini}]{
		\includegraphics{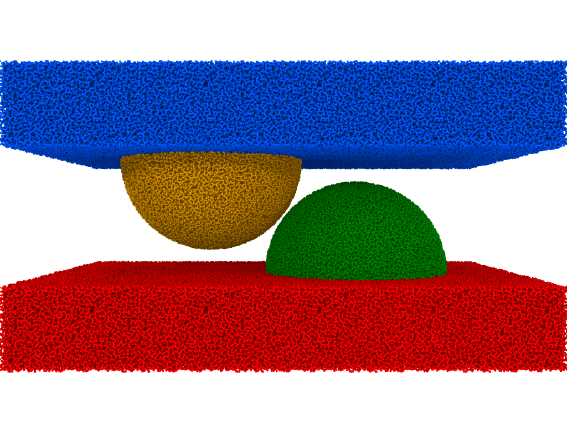}
	}
	\subfloat[After collision\label{subfig:md_20_fin}]{
		\includegraphics{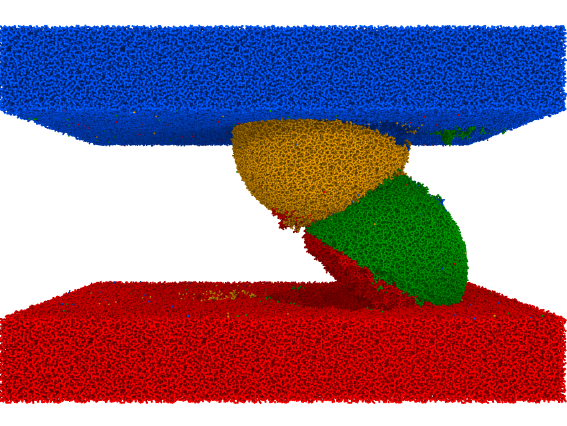}
	}
	\subfloat[Shear strain (cross section)\label{subfig:md_20_def}]{
		\includegraphics{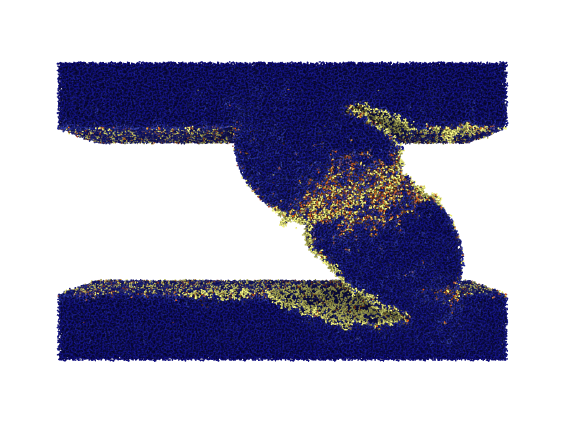}
	}
	\caption[MD simulation of a collision between two amorphous SiO$_2$ asperities of diameter $d = \SI{20}{nm}$]{MD simulation of a collision between two amorphous SiO$_2$ asperities of diameter $d = \SI{20}{nm}$. The initial simulation size is $\SI{50.5}{nm} \times \SI{31.6}{nm} \times \SI{27.5}{nm}$. The presence of clear cracks (regions of high strain) indicates a brittle behavior, implying that $d > d^*$.}
	\label{fig:md_20}
\end{figure*}

To confirm that the estimated $d^*$ really corresponds to the transition of interest for amorphous SiO$_2$, we simulated a setup equivalent to Aghababaei et al. using the Vashishta potential in LAMMPS\cite{plimptonFastParallelAlgorithms1995}. The setup consists of two hemispherical asperities of diameter $d$, each attached to a flat body (as shown in Figure~\ref{fig:md_07}\subref{subfig:md_07_ini}). The amorphous SiO$_2$ structure is created following the same procedure as before. A periodic cubic unit cell of side $\approx\SI{10}{nm}$ containing amorphous SiO$_2$ was generated and used to build the whole setup by tiling the space with the SiO$_2$ cells and carving the desired shapes by removing the atoms, while making sure to preserve charge neutrality. A constant lateral velocity of \SI{10}{m\per s} is applied on the upper body to make the asperities collide. Normal loads are applied on the top and bottom bodies to obtain pressures of \SI{100}{MPa} to \SI{200}{MPa} on the loaded surfaces, which prevents the system from flying apart. Periodic boundary conditions are applied in the horizontal directions. Langevin thermostats are applied at the non-periodic boundaries to keep the temperature of the system constant at \SI{300}{K} with a damping constant of \SI{0.01}{ps}. A timestep of \SI{1}{fs} is used. The simulation results are shown in the Figures \ref{fig:md_07} to \ref{fig:md_20}, visualized using OVITO\cite{stukowskiVisualizationAnalysisAtomistic2009}, for the asperity diameters $d = \SI{7}{nm}$, $d = \SI{10}{nm}$ and $d = \SI{20}{nm}$. These values were chosen around the theoretical estimate $d^* \approx \SI{15}{nm}$. The atomic shear strain (computed relative to the initial configuration) is used to determine whether the wear regime is ductile or brittle. In the ductile case (Figure~\ref{fig:md_07}), the colliding asperities are plastically deformed, resulting after the collision in high permanent shear strains between all the atoms located in the plasticized region. In the brittle case (Figure~\ref{fig:md_20}), a wear particle combining the two asperities is formed, and high strains only remain in the fractured regions, whereas inside the wear particle, the strains increase during the collision and return to a low value after the formation of cracks, since the atoms retain their original configuration. According to Aghababaei et al., the critical length scale $d^*$ is defined as the size of the junction between the colliding asperities corresponding to the transition between ductile and brittle behaviors. In our case, instead of measuring the size of the junction, we assume for simplicity that it has a size roughly equal to the diameter of the asperities $d$. The simulations show that the $d^*$ of amorphous SiO$_2$ is indeed located between \SI{7}{nm} and \SI{20}{nm}. Note that there is no sharp transition between the ductile and brittle behavior (see Figure~\ref{fig:md_10} which shows an in-between behavior), which is why $d^*$ is called a \emph{length scale} rather than a definite length.

The theoretical estimate $d^* \approx \SI{15}{nm}$ and the lower and upper bounds ($d^* > \SI{7}{nm}$ and $d^* < \SI{20}{nm}$) found using simulations for the value of $d^*$ for SiO$_2$ are in line with the minimum size of wear particles found experimentally (no particles smaller than \SI{30}{nm} were observed). This supports the idea that no wear particles smaller than $d^*$ can be created, which is expected because a junction smaller than $d^*$ cannot be detached into a wear particle. Due to the nature of $d^*$ being mainly dependent on material parameters and only weakly on geometry or loading conditions, we can infer that the minimum size of wear particles formed when rubbing two surfaces together is equal to $d^*$ and similarly only dependent on the materials of the two surfaces.

Having established the notion of minimum wear particle size, we know that wear particles can be formed when two asperities located on rubbed surfaces and having a diameter greater than $d^*$ collide into each other. Therefore, wear particles can understandably form from rough surfaces. The formation of wear particles is still possible if the surfaces are flat and free of protruding asperities. When put in contact under a normal load, the surfaces will meet at some points and create adhesive junctions. A tangential load applied on an adhesive junction will enable the formation of wear particles if the size of the junction is greater than $d^*$\snspace. If it is smaller, the junction may grow in size due to plastic deformations, if the adhesive forces are strong enough, until reaching the critical size when a wear particle can finally be detached.

\subsection{Roughness formation}

\begin{figure*}[t]
	\centering
	\subfloat[\SI{1}{s}: formation of first wear particles\label{fig:sem_evo_1s}]{
		\includegraphics{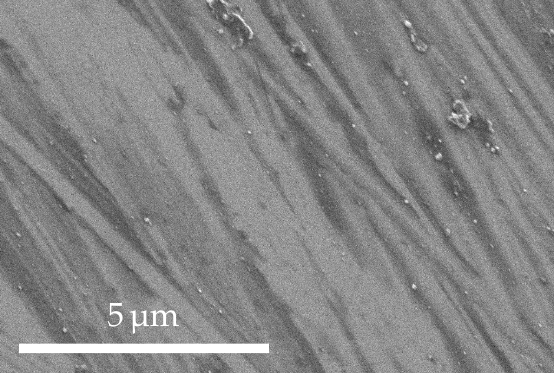}
	}
	\hspace*{1mm}
	\subfloat[\SI{30}{s}: cylinders and aggregates\label{fig:sem_evo_30s}]{
		\includegraphics{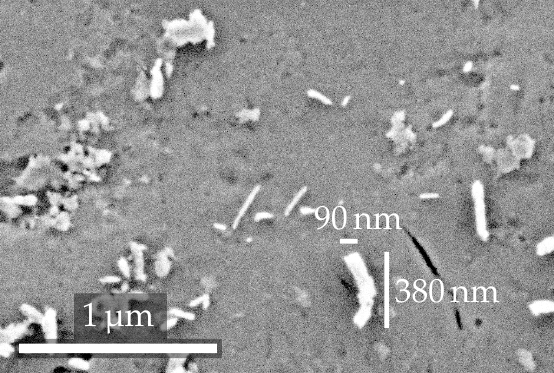}
	}
	\hspace*{1mm}
	\subfloat[\SI{1}{h}: large flakes on the disc forming the wear track\label{fig:sem_evo_1h_a}]{
		\includegraphics{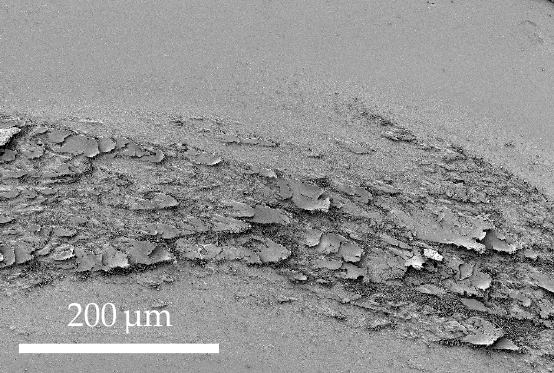}
	}
	\\
	\subfloat[\SI{1}{h}: detail: aggregates forming a large flake\label{fig:sem_evo_1h_b}]{
		\includegraphics{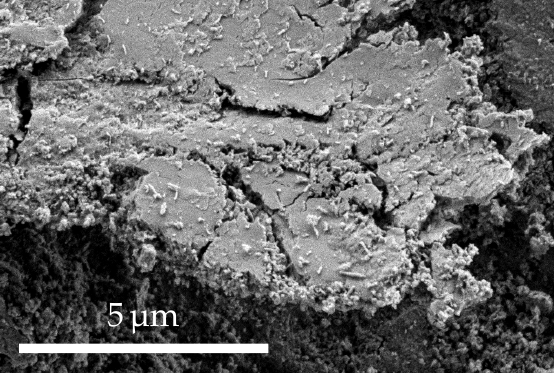}
	}
	\hspace*{1mm}
	\subfloat[\SI{1}{h}: detail: cylinders and aggregates on top of a flake\label{fig:sem_evo_1h_c}]{
		\includegraphics{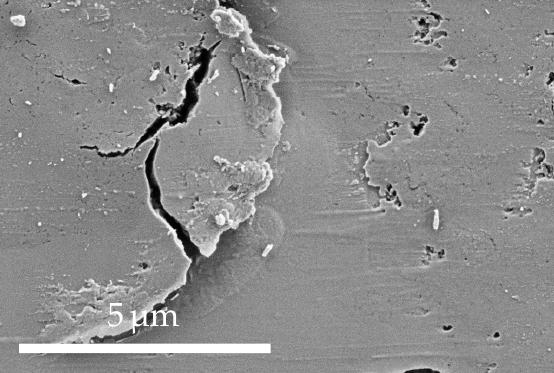}
	}
	\hspace*{1mm}
	\subfloat[\SI{1}{h}: detail: aggregated wear particle fallen outside the wear track\label{fig:sem_aggregate}]{
		\includegraphics{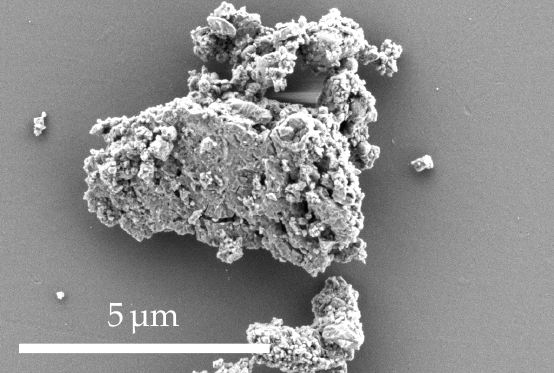}
	}
	\caption{SEM images of wear tracks on flat discs after different amounts of time}
	\label{fig:sem_evo}
\end{figure*}

Figure~\ref{fig:sem_evo} shows SEM images of the wear track left by a ball on flat discs after various sliding durations in chronological order. From an initially almost flat surface of the disc, the process of creation of surface roughness goes through different steps. At first, small spherical wear particles of minimum size $d^*$ are created in the way described in the previous section (see Figure~\ref{fig:sem_1s} and \ref{fig:sem_evo}\subref{fig:sem_evo_1s}). The newly created particles enter a rolling motion between the two rubbed surfaces and start growing in size by chipping away some material from the flat surfaces thanks to adhesive forces (larger particles in Figure~\ref{fig:sem_1s}). The increase of the diameter of the rolling spherical particles is limited by the normal pressure applied by the opposed surfaces, so that the spherical particles will instead continue to grow by elongating into rolling cylinders. With the current loading conditions, the cylinders reach a maximum length of about \SI{500}{nm} and a maximum diameter of \SI{100}{nm} (see Figure~\ref{fig:sem_evo}\subref{fig:sem_evo_30s}). When the cylinders become numerous and meet each other, they agglomerate into larger aggregates. The accumulation of aggregates takes the form of large flakes being left on the initially flat disc, creating a third-body or gouge layer and resulting in the macroscopic surface roughness. The flakes have an average width of \SI{20}{\micro\m} (see Figure~\ref{fig:sem_evo}\subref{fig:sem_evo_1h_a}). The same mechanisms of wear formation continue to take place on top of the already formed third-body layer, as seen in Figures~\ref{fig:sem_evo}\subref{fig:sem_evo_1h_b} and \ref{fig:sem_evo}\subref{fig:sem_evo_1h_c}, where cylinders are visible on top of the large aggregated flakes.

While the creation of wear particles and their accumulation into a third-body layer is commonly seen in engineering \cite{ajayiMechanismTransferFilm1990,katoTribofilmFormationMild2007,kirkEffectFrequencyBoth2019,kirkDebrisDevelopmentFretting2020} or at geological scales\cite{rechesGougeFormationDynamic2005}, the formation of rolling cylinders is interesting in itself and subject to recent discussions in litterature. Rolling cylinders have been observed in lubricated\cite{vargaHydrodynamicsControlShearinduced2019} and unlubricated\cite{zanoriaBallonflatReciprocatingSliding1993,zanoriaFormationCylindricalSlidingWear1995} wear. In the latter case, the cylinders are believed to decrease the interfacial tangential stresses. Reches et al\cite{chenPowderRollingMechanism2017}. have conducted rock friction experiments and observed the formation of rolling cylinders along the sliding direction. They believe the occurrence of these wear cylindrical bodies is a mechanism that leads to slip weakening, with implications in earthquake physics.

\subsection{Roughness evolution}

\begin{figure*}[p]
	\centering
	\begin{tabular}{ccc}
		&
		\subfloat[Initial surface with $Sa = \SI{0.63}{\micro\m}$\label{subfig:surf_05_ini}]{
			\hspace{-12mm}\begin{adjustbox}{clip, trim=0 0 17mm 0}
				\includegraphics{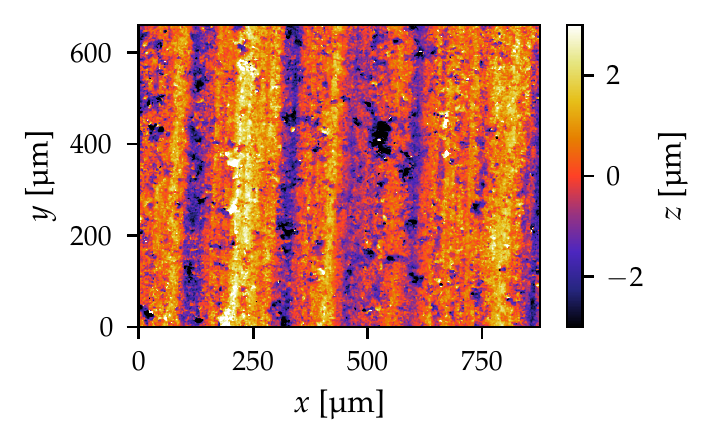}
			\end{adjustbox}
		}
		&
		\subfloat[Initial surface with $Sa = \SI{1.58}{\micro\m}$\label{subfig:surf_10_ini}]{
			\begin{adjustbox}{clip, trim=12mm 0 0 0}
				\includegraphics{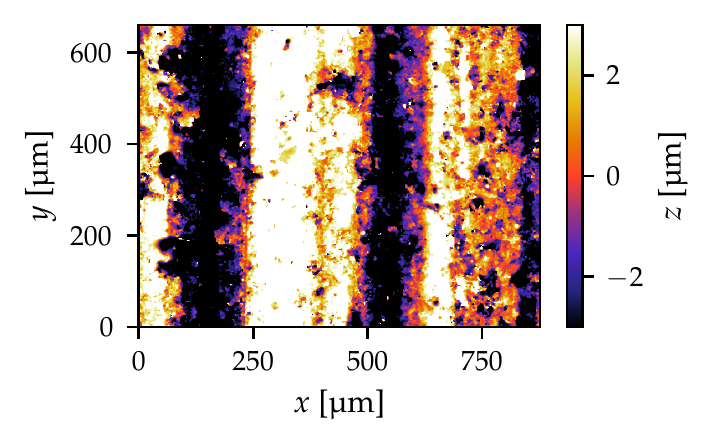}
			\end{adjustbox}
		}
		\\
		\subfloat[Wear track from an initial surface with $Sa \approx \SI{0.01}{\micro\m}$ (not shown). $Sa = \SI{0.23}{\micro\m}$ inside the wear track.\label{subfig:surf_001}]{
			\begin{adjustbox}{clip, trim=0 0 17mm 0}
				\includegraphics{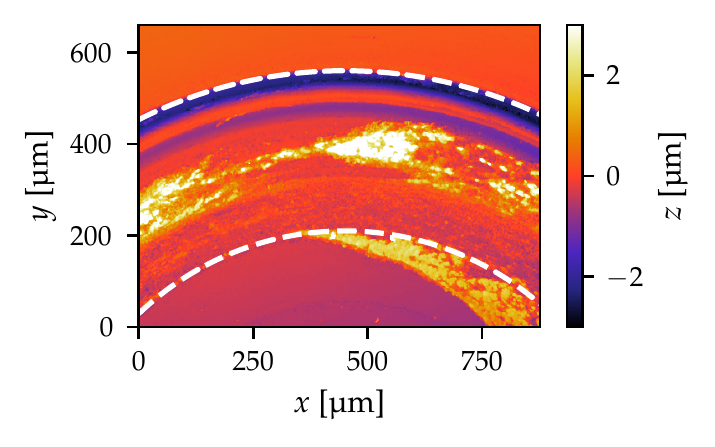}
			\end{adjustbox}
		}
		&
		\subfloat[Wear track from an initial surface with $Sa = \SI{0.63}{\micro\m}$. $Sa = \SI{0.61}{\micro\m}$ inside the wear track.\label{subfig:surf_05}]{
			\begin{adjustbox}{clip, trim=12mm 0 17mm 0}
				\includegraphics{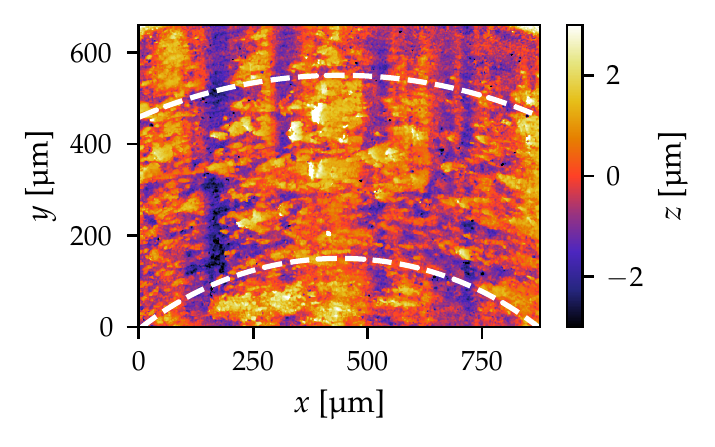}
			\end{adjustbox}
		}
		&
		\subfloat[Wear track from an initial surface with $Sa = \SI{1.58}{\micro\m}$. $Sa = \SI{0.64}{\micro\m}$ inside the wear track.\label{subfig:surf_10}]{
			\begin{adjustbox}{clip, trim=12mm 0 0 0}
				\includegraphics{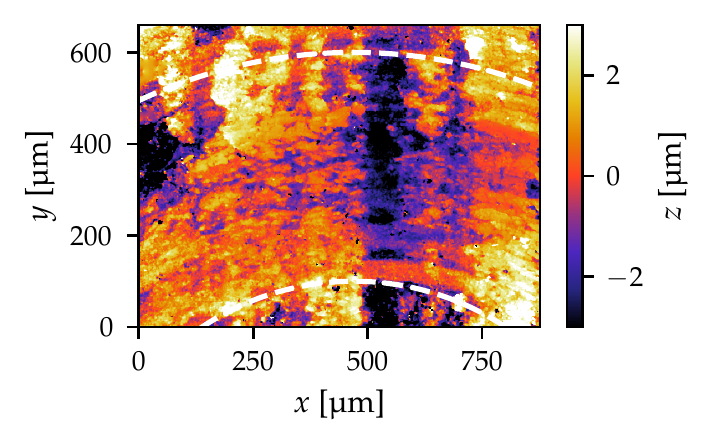}
			\end{adjustbox}
		}
	\end{tabular}
	\caption[Topographic images of discs' initial surfaces and wear tracks after \SI{5}{h} of sliding]{Topographic images of discs' initial surfaces and wear tracks after \SI{5}{h} of sliding. In \textbf{(a)} and \textbf{(b)}, the captions indicate the $Sa$ roughnesses when measured on a window of size $\SI{117}{\micro\m} \times \SI{88}{\micro\m}$ like in the rest of the text.  The portions of circular wear tracks shown in \textbf{(c)}, \textbf{(d)} and \textbf{(e)} have a radius of \SI{1}{mm}. Note that the pairs of images \textbf{(a)}-\textbf{(d)} and \textbf{(b)}-\textbf{(e)} were taken on the same discs, before and after the experiment, but not at the exact same location. The wear track is clearly visible on \textbf{(c)} because it is rougher than the initial surface surrounding it (completely smooth at this scale). The situation is reversed in \textbf{(e)}: the wear track exhibits less dramatic heights than the rougher initial surface. In \textbf{(d)}, the surface roughnesses are similar and the wear track boundaries are harder to distinguish. Note that in \textbf{(c)}, the value of $Sa$ roughness may look smaller than expected. Actually, the roughness averaged over four locations on the wear track is $Sa = \SI{0.77}{\micro\m}$.}
	\label{fig:surfaces}
\end{figure*}

Figure~\ref{fig:surfaces} shows topographic images of disc surfaces with different initial roughnesses (one in each category of Table~\ref{tab:exp_rough}) acquired using a \emph{Sensofar S-Neox} confocal microscope, before and after an experiment with a sliding duration of \SI{5}{h}. Initially, the roughened samples exhibit strong rough features, also revealing the direction of polishing. After the experiments, a wear track is left on each disc, which is visible in the topographic images as well as to the naked eye. When the disc is initially flat, the resulting wear track is comparatively rougher (Figure~\ref{fig:surfaces}\subref{subfig:surf_001}). Conversely, when the disc is initially very rough, the roughness inside the wear track is decreased by the flattening of the high features (Figure~\ref{fig:surfaces}\subref{subfig:surf_10}). In-between, when starting from a relatively moderate surface roughness, the created wear track keeps the same roughness, resulting in a topographic image with uniform heights (Figure~\ref{fig:surfaces}\subref{subfig:surf_05}).

\begin{figure*}[p] % from 09 - Experimental/2019-09-20 - Tests disques SiO2 rugueux/Results.ipynb
	\centering
	\makebox[\textwidth][c]{
		\subfloat[Friction coefficient\label{subfig:evo_rough_mu}]{
			\includegraphics{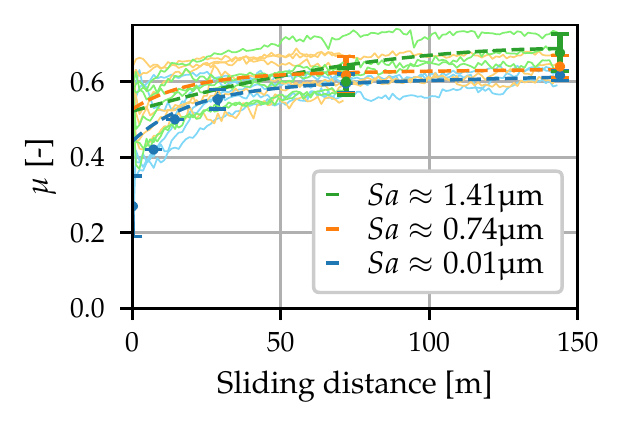}
		}
		\subfloat[Roughness of the wear tracks on the discs\label{subfig:evo_rough_Sa}]{
			\includegraphics{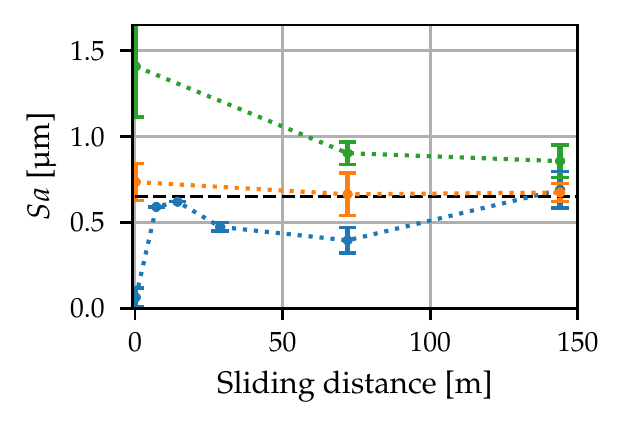}
		}
	}
	\caption[Evolution of friction coefficient and surface roughness from initially smooth and rough SiO$_2$ surfaces]{Evolution of friction coefficient and surface roughness from initially smooth and rough SiO$_2$ surfaces. Errors bars show the standard deviation. \textbf{(a)}~Empirical functions of the form $a + bx + ce^{dx}$ (thick dashed lines) are fitted to the measurements (all thiner curves) for visual comparison. \textbf{(b)}~The roughnesses converge toward the black dashed line placed at $Sa = \SI{0.65}{\micro\m}$. Dotted lines are added to guide the eye and are not representative of the actual evolution paths. The $Sa$ roughnesses are computed from $\SI{117}{\micro\m} \times \SI{88}{\micro\m}$ topographic images of $\SI{1360}{px} \times \SI{1024}{px}$ (details in Appendix~\ref{apx:Sa}).}
	\label{fig:evo_rough}
\end{figure*}

This observed trend in the evolution of surface roughness can be complemented by more accurate topographic computations. Figure~\ref{fig:evo_rough} shows measurements of friction coefficient and of the two-dimensional $Sa$ roughness of the wear tracks left on the discs for all the tests listed in the Table~\ref{tab:exp_rough}.

\begin{figure*}[t!] % from 09 - Experimental/2019-09-20 - Tests disques SiO2 rugueux/Results.ipynb
	\centering
	\makebox[\textwidth][c]{
		\subfloat[Example of computation of Hurst exponent on a disc with an initial roughness of $Sa = \SI{0.79}{\micro\m}$ after \SI{5}{h} (\SI{144}{m}) of sliding\label{subfig:H_comp}]{
			\includegraphics{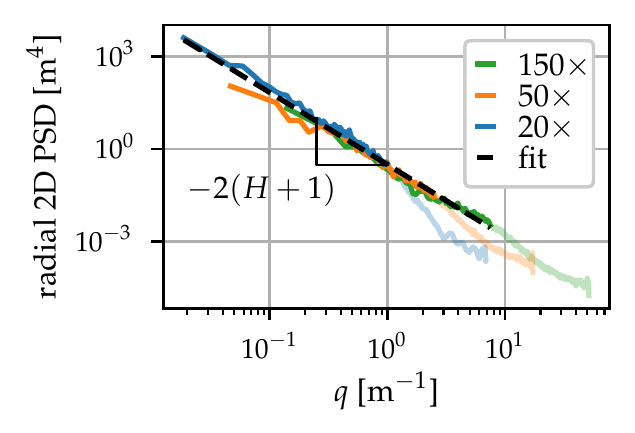}
		}
		\subfloat[Hurst exponent of the wear tracks on the discs\label{subfig:evo_hurst_H}]{
			\includegraphics{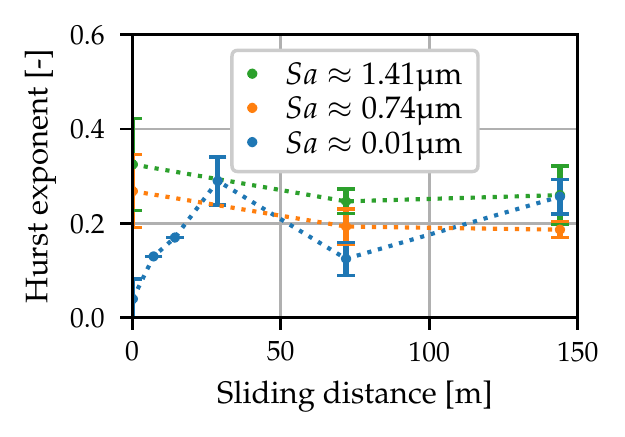}
		}
	}
	\caption[Example of computation of the Hurst exponent and its evolution from initially smooth and rough SiO$_2$ surfaces]{Example of computation of the Hurst exponent and its evolution from initially smooth and rough SiO$_2$ surfaces. \textbf{(a)}~Three radial PSDs are obtained using three different magnifications on a sample, centered on the same spot. The black dashed line is fitted to all the three curves, leading to the Hurst exponent $H = 0.17$ in this case. The transparent parts of the curves are above the cutoff frequency and do not contribute to the fit. The PSD obtained from the lowest magnification (20$\times$) gives low frequency informations while the one obtained from the highest (150$\times$) provides high frequencies. More details are given in Appendix~\ref{apx:hurst}. Numerical data for the three PSD curves shown here and for other cases can be found in Supplementary material \hyperref[suppl]{S1}. \textbf{(b)}~The Hurst exponents start scattered and evolve toward a similar value. Errors bars show the standard deviation. Dotted lines are added to guide the eye and are not representative of the actual evolution paths.}
	\label{fig:evo_hurst}
\end{figure*}

The evolution of the friction coefficients indicates that an almost steady state is reached after a running-in period. The running-in period itself does not seem to last a constant time between the different tests. For example, the wear track on the disc can start to become visible to the naked eye on the initially flat discs after 30 seconds or only after 5 seconds of sliding for two repetitions of the same test. The running-in period may be very dependent on small non-controlled variations of the initial conditions (like dust, temperature or hygrometry). The friction coefficient still increases slowly in the `steady' region, which is probably linked to the non-constant contact area between the worn ball and the disc, which increases as the ball wears out.

The dashed curves of the Figure~\ref{fig:evo_rough}\subref{subfig:evo_rough_mu} are fitted to all the evolution curves of a given category of initial surface roughness of the disc (corresponding to the three categories of the Table~\ref{tab:exp_rough}). Functions of the form $a + bx + ce^{dx}$ are empirically chosen, only for visual comparison purposes, as they can represent a steadily increasing regime with the $a + bx$ terms and an exponential convergence to this regime with the $ce^{dx}$ term. The friction coefficient is in average higher when the initial surface roughness is higher. However, the fitted curves are within a standard deviation apart from each other, so this trend is not statistically significant.

In contrast, the evolution of the surface roughness of the discs over time has an interesting dependence on the initial disc roughness: when the initial roughness is in the explored range (from $Sa = \SI{0.01}{\micro\m}$ to $Sa = \SI{1.58}{\micro\m}$), the roughness evolves toward a common value of $Sa \approx \SI{0.65}{\micro\m}$ (the black dashed line in Figure~\ref{fig:evo_rough}\subref{subfig:evo_rough_Sa}) irrespective of the initial value.

Note that the average values of initial roughnesses of $Sa \approx \SI{0.74}{\micro\m}$ and $Sa \approx \SI{1.41}{\micro\m}$ were chosen after witnessing that initially almost flat discs evolved toward a roughness of $Sa \approx \SI{0.65}{\micro\m}$. The grain sizes of the sandpapers used in the creation of the rough discs were selected to obtain a roughness being around 1 time and 2 times the value $Sa \approx \SI{0.65}{\micro\m}$ after polishing.

While convenient at larger scales, one drawback of the $Sa$ roughness measurement is its dependence to the scale of measurement for fractal surfaces. Typically, the measured $Sa$ roughness is lower at smaller scales. Instead, the computation of the power spectral density (PSD) of the surface heights\footnote{It is defined as the squared norm of the two-dimensional Fourier transform of the surface heights. See Appendix~\ref{apx:hurst} for details.} (one example shown in Figure~\ref{fig:evo_hurst}\subref{subfig:H_comp}) shows that all PSD curves follow a power-law over several orders of magnitude, which is typical of self-affine surfaces\cite{perssonContactMechanicsRandomly2006}, meaning that all initial and worn surfaces in our experiments are remarkably self-affine. Our results reveal that the resulting surfaces are self-affine on over two orders of magnitude (Figure~\ref{fig:evo_hurst}\subref{subfig:H_comp}). For such self-affine surfaces, another measure of roughness besides the $Sa$ roughness is the Hurst exponent, which is independent of the measuring scale and is related to the slope of the PSD in a logarithmic plot\cite{jacobsQuantitativeCharacterizationSurface2017}. The PSD $C(q)$ of a self-affine surface takes the form
\begin{equation}
    C(q) \propto q^{-2(H+1)}
\end{equation}
in a certain range, where $q$ is the radial spatial frequency and $H$ is the Hurst exponent. The Hurst exponent of a self-affine surface is related to its fractal dimension, and it describes how the roughness changes when viewing the surface from a larger or smaller perspective\cite{mandelbrotFractalCharacterFracture1984}. Figure~\ref{fig:evo_hurst}\subref{subfig:evo_hurst_H} shows measurements of the Hurst exponent of the wear tracks left on the discs for the tests listed in the Table~\ref{tab:exp_rough}. Remarkably, while the initial Hurst exponents differ depending on the initial roughness, all our measured Hurst exponents converge toward a common value of $H \approx 0.25$.

\begin{figure}
	\centering
	\includegraphics{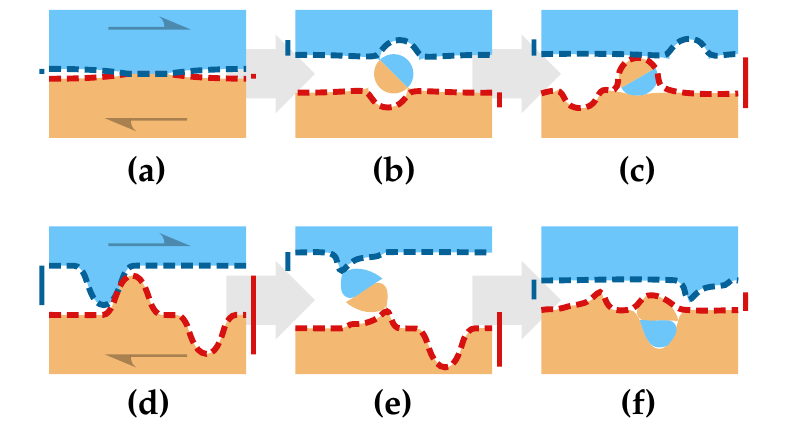}
	\caption[Schematics of roughness evolution behaviors]{Schematics of roughness evolution behaviors. The vertical bars on the sides of each panel represent the total height contributing to surface roughness. \textbf{(a)}~Starting from low initial roughness. An adhesive junction forms between the two surfaces. \textbf{(b)}~Formation of a wear particle. The roughness increases with the newly formed cracks. \textbf{(c)}~The particle attaches itself to one surface, increasing again the roughness. \textbf{(d)}~Starting from high initial roughness. \textbf{(e)}~Formation of a wear particle from the colliding asperities. The roughness decreases with their destruction. \textbf{(f)}~The particle falls into a hole, decreasing again the roughness.}
	\label{fig:schem_rough_evo}
\end{figure}

The different behaviors of surface roughness evolution can be qualitatively explained with a simple physical intuition like shown in Figure~\ref{fig:schem_rough_evo}. When the surface roughness is low, wear particles can be chipped away from the contacting surfaces thanks to adhesive forces, as described in Section~\ref{sec:ini_particle}. The cracks created by the detachment of particles contribute to increase the surface roughness. Conversely, when the surface roughness is large, high peaks on the surfaces can meet and knock each other out, and created wear particles can fill up holes in the surfaces, reducing the surface roughness. The two mechanisms continuously happen simultaneously over the contacting surfaces, balancing each other. Thus, a stable surface roughness can be reached after a running-in period. Nonetheless, this simple description does not explain the remarkable finding that the final roughness profile does not seem to retain any memory of the initial roughness.

Note that while the surface roughness of the worn parts obtained after a running-in period appear to be independent of their initial surface roughness, the final steady-state friction coefficient and surface roughness are likely dependent on the loading parameters (normal load, turning radius and sliding velocity) and the material properties. Thus, the particular value of $Sa \approx \SI{0.65}{\micro\m}$ is specific to the current set of parameters. A change in those parameters might change the target value of $Sa$, a behavior which has not been studied here.

All of our analyses on roughness evolution and formation were performed on the discs and not on the spherical pins. The loading conditions (\emph{pin-on-disc}) imply that the pins are more locally solicited than the discs, resulting in an asymmetric evolution of the worn surfaces. Overall, the worn caps of the balls have a $Sa$ surface roughness two orders of magnitude lower than their disc counterparts, so that the most interesting features created by wear could only be found on the discs. The third-body layer is indeed only deposited on the discs, creating the macroscopic roughness.

\section{Conclusion}

We performed tribological experiments on amorphous SiO$_2$ samples of different surface roughnesses. We found that for the employed loading conditions (load, sliding velocity), a self-affine surface roughness emerges from initially comparatively flat surfaces via the following steps:
\begin{enumerate}
	\item Formation of small spherical wear particles, whose minimum size is mainly dependent on the material properties. These initial wear particles are the fundamental bricks in the formation of the third-body layer;
	\item The wear particles grow into rolling cylinders and into larger aggregates;
	\item The large aggregates take the form of flakes, creating a third-body layer and the macroscopic roughness.
\end{enumerate}
Alternatively, when starting with an initially high surface roughness, the latter gets reduced by wear to a value similar to the one obtained from initially flat surfaces. The initial surface roughness does not influence the state (surface roughness) and behavior (friction coefficient) of the worn parts attained after a running-in period. Thus, a stable value for the coefficient of friction could be achieved by machining an initial surface roughness that matches the expected final roughness, which depend on the materials and loading conditions. Further studies can be considered to explore the effects of the loading conditions and material properties (and therefore the minimum wear particle size) on the steady state values of surface roughness and friction coefficient, if such steady state can be reached, which could be conducted either experimentally or using computer simulations.

\section*{Acknowledgments}

We thank Tobias Brink\footnote{Department of Structures and Nano/Micromechanics of Materials, Max-Planck-Institut für Eisenforschung, 40237 Düsseldorf, Germany} for sharing his expertise in MD simulations and for providing insightful discussions.

% \newpage
\appendix

\section{Appendix}

\subsection{Calculation of contact diameter and pressure}\label{apx:hertz}

The material properties of SiO$_2$ are $E = \SI{73}{GPa}$ and $\nu = 0.17$. The contact radius $a$ is given by\cite{johnsonContactMechanics1985a}:
\begin{equation}
	a = \sqrt[3]{\frac{3F_\text{N}R}{4E^*}} = \SI{39.1}{\micro\m}
\end{equation}
where $R = \SI{3}{mm}$ and $E^*$ is the equivalent Young's modulus:
\begin{equation}
	E^* = \frac{E}{2(1-\nu^2)} = \SI{37.6}{GPa} \,.
\end{equation}

The maximum pressure reached at the center of the contact zone is:
\begin{equation}
	p_0 = \frac{3F_\text{N}}{2\pi a^2} = \SI{312}{MPa} \,.
\end{equation}
For comparison, the compressive strength of SiO$_2$ is $\sigma_\text{m} = \SI{1150}{MPa}$.

\subsection{$Sa$ roughness measurement}\label{apx:Sa}

Surface topography measurements are acquired using a \emph{Sensofar S-Neox} confocal microscope on a $\SI{117}{\micro\m} \times \SI{88}{\micro\m}$ window of $\SI{1360}{px} \times \SI{1024}{px}$, resulting in data points $h(x, y)$ with discrete $x$ and $y$. Missing data points are filled by iteratively solving a Laplace equation at those points:
\begin{equation}
	\frac{\partial^2 h}{\partial x^2} + \frac{\partial^2 h}{\partial y^2} = 0 \,.
\end{equation}
A plane is fitted by least-square minimization and subtracted to the points to obtain a zero-mean topography $h_0(x, y)$. The $Sa$ roughness defined as:
\begin{equation}
	Sa = \frac{1}{N} \sum_{x, y} |h_0(x, y) - \overline{h}_0|
\end{equation}
where $N$ is the number of data points and $\overline{h}_0$ is the mean value of $h_0$.

Since the value of $Sa$ is dependent on the size of the measurement site due to the fractal nature of the rough surfaces, all measurements of $Sa$ roughness have to be performed at the same window size to be comparable. Usually, the $Sa$ roughness decreases when the window size gets smaller.

To obtain the $Sa$ roughness of the unpolished and polished discs before the tests, only one topographic measurement is performed at the center of the sample. For the $Sa$ of the circular wear tracks left on the discs, four locations on the track are measured and the computed $Sa$ are averaged.

\subsection{Hurst exponent measurement}\label{apx:hurst}

The Hurst exponent of a rough surface is computed from topographic measurements. Three measurement are made using the three magnifications 20$\times$, 50$\times$ and 150$\times$, leading to images of $\SI{1360}{px} \times \SI{1024}{px}$ on windows of size $\SI{877}{\micro\m} \times \SI{660}{\micro\m}$, $\SI{351}{\micro\m} \times \SI{264}{\micro\m}$ and $\SI{117}{\micro\m} \times \SI{88}{\micro\m}$ respectively. Note that when performing a measurement for a wear track at the lowest magnification (the largest scale), the wear track does not fill the whole image (see Figure~\ref{fig:surfaces}), so it has to be cropped.

For each topographic data $h(x, y)$, a radial Hann window is applied and the discrete Fourier transform $H(q_x, q_y)$ of the windowed data is computed, as well as the power spectral density (PSD):
\begin{equation}
	C(q_x, q_y) = \frac{L_xL_y}{n_x^2n_y^2} |H(q_x, q_y)|^2 \,,
\end{equation}
where $q_x$ and $q_y$ are the spatial frequencies, $L_x$ and $L_y$ are the window size and $n_x$ and $n_y$ are the window resolution. The radial PSD $C(q)$ with $q^2 = q_x^2 + q_y^2$ is computed form the cartesian PSD $C(q_x, q_y)$ by binning into 512 linearly spaced bins. Due to physical limitations in the measuring hardware, higher frequencies of the PSDs are not representative and have to be dropped. A cutoff of \SI{0.05}{m^{-1}} times the magnification (20, 50 or 150) is applied to the radial PSDs. A line is fitted to the three PSDs in a log-log graph, ignoring the values after the cutoff frequencies. The slope of the line is $-2(H+1)$, where $H$ is the searched Hurst exponent.

The Figure~\ref{fig:evo_hurst}\subref{subfig:H_comp} shows an example of Hurst exponent computation from three radial PSDs obtained at the different magnifications. Furthermore, the Supplementary material \hyperref[suppl]{S1} contains datapoints for other radial PSD curves corresponding to other sliding durations and initial surface roughness. Those PSD curves can be used to numerically reconstruct surfaces with similar roughness properties.

\section*{Supplementary material}\label{suppl}

Supplementary material associated with this article can be found along its online version.

\atColsBreak{\vskip10pt}
\printbibliography

\end{document}